\documentstyle[psfig]{l-aa}

\begin{document}

   \thesaurus{07          
              (02.03.3;   
               08.01.1;   
               08.01.3;   
               08.16.3;   
               02.12.1)}  

   \title{
3D hydrodynamical model atmospheres of metal-poor stars}

   \subtitle{Evidence for a low primordial Li abundance}

   \author{Martin Asplund\inst{1}, \AA ke Nordlund\inst{2}, Regner
Trampedach\inst{3}, Robert F. Stein\inst{3}
          }

   \offprints{martin@nordita.dk}

   \institute{NORDITA, 
              Blegdamsvej 17, 
              DK-2100 ~Copenhagen {\O}, 
              Denmark\\ 
              \and
              Astronomical Observatory, NBIfAFG, 
              Juliane Maries Vej 30,
              DK-2100 Copenhagen \O, Denmark\\
              \and
              Department of Physics and Astronomy, 
              Michigan State University, 
              East Lansing, MI 48823, 
              USA
              }

   \date{Received: April 19, 1999; accepted: May 5, 1999}

   \maketitle

   \markboth{Asplund et al.: 
3D hydrodynamical model atmospheres of metal-poor stars
}
{Asplund et al.: 3D hydrodynamical model atmospheres of metal-poor stars
}

\begin{abstract}

Realistic 3-dimensional (3D), radiative hydrodynamical surface
convection simulations of the metal-poor halo stars HD\,140283 and
HD\,84937 have been performed. 
Due to the dominance of adiabatic cooling over radiative heating
very low atmospheric temperatures 
are encountered. The lack of spectral lines in these
metal-poor stars thus causes much steeper temperature gradients
than in classical 1D hydrostatic model atmospheres where the temperature of
the optically thin layers is determined by radiative 
equilibrium.  The modified atmospheric structures cause
changes in the emergent stellar spectra.
In particular, the primordial Li abundances may have been
overestimated by 0.2-0.35\,dex with 1D model atmospheres.
However, we caution that our result assumes
local thermodynamic equilibrium (LTE), while the 
steep temperature gradients may be prone to e.g. over-ionization.

   \end{abstract}

\section{Introduction}

Stellar chemical compositions are of great astrophysical importance
as they carry information on stellar evolution and mixing processes,
galaxy formation and evolution, and
Big Bang nucleosynthesis and the baryon density of the Universe. 
In order to decipher the observed stellar spectra in terms of
abundances a proper understanding of the line formation process
is required. 
Often-used assumptions and approximations in abundance analyses
involve 1D, plane-parallel, hydrostatic and flux-constant 
(radiative and convective equilibrium) model 
atmospheres in LTE. Free parameters
such as the mixing length parameters, micro- and macroturbulence enter 
the model constructions and the spectral syntheses, and
cause uncertainties in the derived abundances.

An improved theoretical foundation for the interpretation
of stellar spectra is desirable, to 
complement the impressive recent observational advances
in terms of signal-to-noise, spectral resolution and limiting magnitudes.
Modern supercomputers now allow self-consistent 3D, radiation-hydrodynamics
simulations of the surface convection of late-type stars.
These simulations invoke no free parameters, 
yet succeed in reproducing key
diagnostics such as granulation topology and statistics, 
helioseismic properties,
and spectral line shapes, shifts and asymmetries for the 
best available test-bench: the Sun (Stein \& Nordlund 1998; 
Rosenthal et al. 1999; Asplund et al. 1999b).
Since convection may be expected to have a prominent influence on 
the atmospheres of metal-poor halo stars we have performed 3D 
simulations for a number of such stars.  In the present {\it Letter}, we 
briefly present results for two of the stars and compare 
emergent spectra from the 3D models with those of conventional 1D 
models.  A full account of these and other metal-poor convection 
simulations will be presented elsewhere.

\section{Hydrodynamical surface convection simulations}

Realistic 3D, time-dependent, surface convection simulations
of the metal-poor halo stars HD\,140283 and HD\,84937 have been performed
using a compressible radiative-hydrodynamical code, 
that has previously been successfully applied to studies 
of the solar granulation (Stein \& Nordlund 1998). 
The equations of mass, momentum and energy conservation coupled
to the 3D equation of radiative transfer are solved on an Eulerian
mesh with 100\,x\,100\,x\,82 zones, covering
35\,x\,35\,x\,12\,Mm (HD\,140283) and 21\,x\,21\,x\,8\,Mm (HD\,84937),
respectively.

In order to accurately describe the photospheric layers, special
care must be taken to include appropriate input physics. 
A state-of-the-art equation-of-state (Mihalas et al. 1988),
which includes the effects of ionization, excitation and dissociation,
has been used together with detailed continuous 
(Gustafsson et al.\ 1975 and subsequent updates) and 
line (Kurucz 1993) opacities. 
The 3D radiative transfer is solved for one vertical and four inclined
rays under the approximations of local thermodynamical equilibrium (LTE)
and grouping of the opacities into four bins (Nordlund 1982). 
The accuracy of the opacity binning procedure is verified
throughout the simulations by solving the full monochromatic
radiative transfer (2748 wavelength points) in the 1.5D approximation
and found to always agree within 1\% in emergent flux. 
It is noteworthy that the convection simulations 
contain no adjustable parameters 
besides those specifying the effective temperature $T_{\rm eff}$ 
(or, as used here, the entropy of the
inflowing gas at the lower boundary), 
the gravitational
acceleration at the surface log\,$g$ and the chemical composition [Fe/H].
The adopted stellar parameters, which have been obtained 
from the infrared flux method (IRFM, Alonso et al. 1996)
for $T_{\rm eff}$, Hipparcos parallaxes for log\,$g$ and published
values for [Fe/H], are listed
in Table \ref{t:param}. The individual elemental abundances are taken from 
Grevesse \& Sauval (1998) scaled appropriately.
Further details on the analogous solar simulations 
may be found in Stein \& Nordlund (1998).

\begin{table}[t]
\caption{Adopted stellar parameters 
\label{t:param}}
\begin{tabular}{lcccc}
\\ 
\hline \\
Star       & $T_{\rm eff}$ & log\,$g$ & [Fe/H] & $\xi_{\rm turb}^a$ \\
           & [K]           & [m s$^{-2}$]    &        & [km\,s$^{-1}$]\\
\hline \\
HD\,140283 & 5690          &  1.67    &  -2.50 & 1.0 \\
HD\,84937  & 6330          &  2.04    &  -2.25 & 1.0 \\
\hline  \\
\end{tabular}
\begin{list}{}{}
\item[$^{\rm a}$] Not necessary for the 3D spectral line synthesis 
\end{list}
\end{table}

\begin{figure}[t]
\centerline{
\psfig{figure=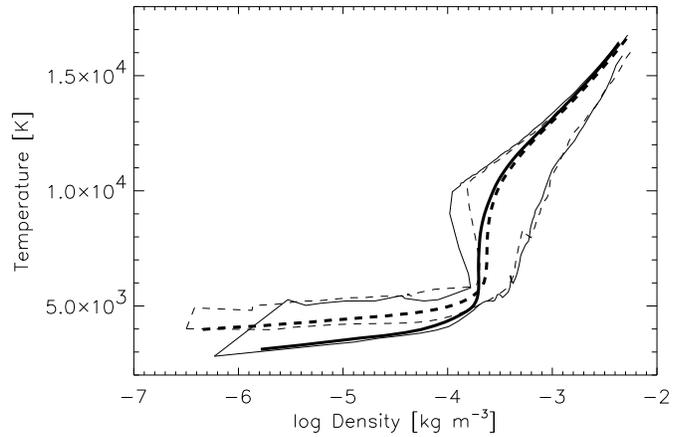,width=9.cm}
}
\caption{Average T($\rho$) for HD\,140283 (solid) compared with the Sun 
(dashed). Thin solid and dashed curves represent the
corresponding extreme temperatures for a given density 
}
         \label{f:trho}
\end{figure}

\begin{figure}[t]
\centerline{
\psfig{figure=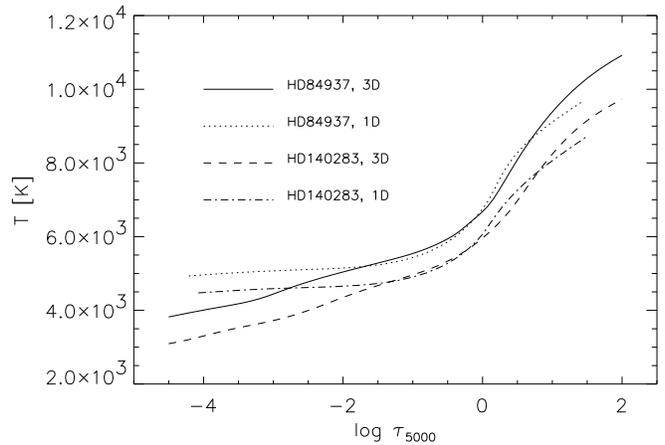,width=9.cm}
}
\caption{$T(\tau)$ for 3D and 1D models of HD\,140283 and HD\,84937
}
         \label{f:ttau}
\end{figure}

Fig. \ref{f:trho} shows the resulting temperature structure for
the HD\,140283 simulation, together with the corresponding 
solar simulation (Asplund et al. 1999b). 
The similarity between the two simulations in the deeper layers is
partly fortuitous, and is due to an almost perfect 
cancellation of the effects
of different log\,$g$ and [Fe/H]. More striking are the prominent
differences in the temperature structure of the optically thin layers. 
While the solar simulation remains close to
radiative equilibrium, much lower temperatures are encountered in
the HD\,140283 simulation. This seems to be a generic feature 
as is also present in our
simulations of HD\,84937 and other low-metallicity stars.
The large differences also remain when comparing
on optical depth scales as done in Fig. \ref{f:ttau} where the
spatial (on surfaces of equal optical depths) and temporal averages from the
3D simulations are presented together with the corresponding 
1D {\sc marcs} structures. 

The temperatures in the optically thin layers are predominantly determined
by two competing effects: adiabatic cooling due to the expansion of
ascending material and radiative heating by spectral lines
(radiative heating occurs when spectral lines reabsorb some of
the continuum photons whose release cause the intense cooling near 
continuum optical depth unity).
Thus, in dynamical atmospheres the temperature tends to become depressed
below its radiative equilibrium value and lines act mainly as heating agents.
This is opposite to the case when radiative equilibrium is enforced
in theoretical 1D model atmospheres.  There,
spectral lines cause surface cooling (e.g.\ Mihalas 1978), hence leading
to shallower temperature gradients when the metallicity is diminished.
In the convection simulations, the opposite happens: when 
decreasing [Fe/H], fewer and weaker lines are available 
to absorb photons, allowing the adiabatic cooling to dominate more.
As a result, balance between heating and cooling is achieved 
at lower surface temperatures, causing a steeper $T(\tau)$.

\section{Impact on spectral line formation}

Using the convection simulations as model atmospheres the 3D radiative
transfer was computed under the assumption of LTE  
for lines of special astrophysical
interest, as listed in Table \ref{t:abund}.
From the full simulations representative sequences
of one hour with snapshots every 2 min 
were selected for the line calculations;
the emergent fluxes for the shorter intervals correspond to 
$T_{\rm eff} = 5672\pm3$\,K (HD\,140283) and
$T_{\rm eff} = 6356\pm8$\,K (HD\,84937), respectively, i.e. very
close to the intended $T_{\rm eff}$.
Prior to the spectral synthesis the simulations were interpolated
to a finer vertical resolution ($\tau_{\rm max} = 300$).
The radiative transfer equation was solved for a total of 29 directions 
(4 $\mu$-angles and 7 azimuthal angles plus the vertical)
after which a disk-integration was carried out. 
Elemental abundances were derived from equivalent widths, to allow direct 
comparison with 1D models.
It should be noted that no micro- or macroturbulence enter
the calculations, as the convective Doppler shifts and
velocity gradients 
are sufficient to produce the observed line broadening  
(Asplund et al. 1999b).
To estimate the impact of 3D models, a differential comparison
with classical 1D {\sc marcs} model atmospheres 
(Gustafsson et al.\ 1975 with subsequent updates) was carried
out, adopting a microturbulence of 1.0\,km\,s$^{-1}$.
The stellar parameters for the 1D models are identical to those
for the shorter convection simulation sequences.

The reliability of the simulations 
may be assessed from a comparison between observed
and predicted line asymmetries. For HD\,140283 the theoretical
bisectors agree very well with observations 
(Allende Prieto et al. 1999; Asplund et al. 1999a), 
e.g.\ in explaining the greater line asymmetries
in this metal-poor star compared with solar metallicity 
comparison stars.
The good concordance provide additional confidence in
the simulations. 
 
\begin{table}[t]
\caption{The derived LTE abundances of 3D hydrodynamical simulations
compared with 1D hydrostatic model atmospheres (on the customary logarithmic 
scale where log\,$\epsilon_{\rm H}$=12.00)
\label{t:abund}}
\begin{tabular}{lcccc} 
\\
\hline \\
Line & \multicolumn{2}{c}{HD\,140283} & \multicolumn{2}{c}{HD\,84937} \\
$$[nm] & 3D  & 1D$^a$ & 3D & 1D$^a$ \\
\hline \\
Li\,{\sc i}\,670.8  & $1.78$ & $2.12$ & $2.08$ & $2.28$ \\
Be\,{\sc ii}\,313.1 &$-0.92$ &$-1.06$ &$-0.85$ & $-0.88$ \\
B\,{\sc i}\,249.8   &$-0.41$ &$-0.21$ &        &  \\
O\,{\sc i}\,777.2   & $7.20$ & $7.15$ & $7.32$ & $7.30$ \\
K\,{\sc i}\,769.9   & $2.60$ & $2.74$ &        &  \\
Ca\,{\sc i}\,616.2  & $3.62$ & $3.77$ & $4.22$ & $4.35$ \\
Fe\,{\sc i}$^b$          & $4.57\,(0.16)$ & $5.02\,(0.17)$ & $5.10\,(0.12)$ & $5.37\,(0.11)$ \\
Fe\,{\sc ii}$^b$         & $5.16\,(0.10)$ & $5.08\,(0.11)$ & $5.38\,(0.02)$ & $5.35\,(0.03)$ \\
Ba\,{\sc ii}\,614.2 &$-1.28$ &$-1.12$ &$-0.22$ & $-0.05$ \\
\hline  \\
\end{tabular}

\begin{list}{}{}
\item[$^{\rm a}$] Adopting $\xi_{\rm turb} = 1.0\,$km\,s$^{-1}$
\item[$^{\rm b}$] 10 Fe\,{\sc i} lines, 2 Fe\,{\sc ii} lines. 
The dispersions are in parenthesis

\end{list}

\end{table}

Due to differences in their sensitivity to the temperature
structure, lines are influenced differently by the differences
between 3D convection simulations and 1D model atmospheres, as is
evident from Table \ref{t:abund}.
Lines of neutral minority species, low excitation transitions
and stronger lines are formed in higher layers and 
thus feel the low temperatures there, 
making them stronger for a given abundance. 
We note for example that from the resonance line of Li\,{\sc i} a
0.2-0.35\,dex lower abundance is derived from the 3D models than
from the 1D models. 
On the other hand, high excitation transitions and lines of
ionized elements tend to result in slightly larger abundances
for the 3D models,
as is the case for Be\,{\sc ii}, O\,{\sc i} and Fe\,{\sc ii}.
The larger effects for the HD\,140283 simulation compared
with HD\,84937 may be attributed to the generally
stronger lines in the former due to its lower $T_{\rm eff}$.
Finally, we stress that the main lesson to draw from 
Table \ref{t:abund} is not the {\em absolute} 3D abundances 
but the {\em relative} abundances compared with 
differential 1D model atmospheres.

\section{Discussion}

It is noteworthy that ionization balance of Fe is not fulfilled
for the two 3D simulations, in contrast to the 1D {\sc marcs} 
models, which suggests that either 
1) the adopted stellar parameters are inappropriate,
2) the atmospheric temperatures have been underestimated, 
or 3) there are significant departures from LTE present in the
form of a general over-ionization of Fe of 0.3-0.6\,dex. 
The latter effect is certainly not inconceivable, given the 
steep $T(\tau)$ and the weak UV-blocking in these metal-poor
environments; a 3D non-LTE investigation is
clearly of high priority. 
One may speculate that if indeed Fe is over-ionized compared
with LTE expectations, Li may also show a similar effect, in 
spite of the generally small non-LTE corrections in the 1D case
(Carlsson et al. 1994). 
The possibility of erroneous $T_{\rm eff}$s can most likely be refuted,
since preliminary calculations reveal only minor differences
 ($\la 50\,$K) in IRFM estimates between our 3D simulations and the
1D {\sc marcs} models.  
The uncertainties in the stellar parameters 
($\Delta T_{\rm eff} = 80$\,K, $\Delta$log\,$g = 0.15$) 
only correspond to $\Delta$[Fe\,{\sc i}]\,$= 0.10$ and
$\Delta$[Fe\,{\sc ii}]\,$= 0.08$ and can thus 
probably not be blamed for the discrepancy either.

One may suspect that the very low surface temperatures have
indeed been underestimated, causing systematic
effects on the derived abundances.
Support for such a suspicion comes from an observed trend
between abundance and excitation potential for the 
Fe\,{\sc i} lines, that is present with both metal-poor
simulations but slightly less pronounced and in opposite direction 
with the 1D models; unfortunately
the sample of lines is still somewhat limited however. 
As discussed above, it is physically very plausible
to have sub-radiative equilibrium temperatures in these
metal-poor environments, but the magnitude of the effect
may have been overestimated.
One possible reason could be the neglect of Doppler
shifts in the radiative energy transfer employed in 
the convection simulations.  
In the optically thin layers the temperature is largely controlled 
by only a few strong lines that are partially saturated.
Velocity gradients will enable the lines to absorb
unattenuated continuum photons instead, thereby causing additional
heating. Improved simulations addressing also this remaining 
refinement is planned.
We note, however, that our solar simulations already predict
essentially perfect agreement of line shapes and asymmetries 
with observations (Asplund et al.\ 1999b), as well as agreement
for a range of other diagnostics.

Though the results presented here should be considered 
preliminary, pending further investigations of
departures from LTE and the reality of the low
surface temperature,
it is still interesting to estimate possible effects on  
Big Bang nucleosynthesis and galactic chemical evolution.

Our 3D results suggest that the primordial Li abundance
may have been overestimated by $\approx 0.25$\,dex when
using classical 1D model atmospheres. Adopting 
log$\,\epsilon_{\rm Li} = 2.20 \pm 0.03$
as the best estimate of the Li-plateau from a 
sample of metal-poor halo stars on the same
$T_{\rm eff}$-scale as used here (Bonafacio \& Molaro 1997),
we conclude that the primordial Li abundance may be as low as
log$\,\epsilon_{\rm Li} = 1.95 \pm 0.03$, 
corresponding to Li/H=$8.9\pm0.6\cdot10^{-11}$.
It should be noted that the $1\sigma$ uncertainties 
only represent the (1D) statistical errors
(Ryan et al. 1999) and no corrections for departures from LTE, 
post-Big Bang nucleosynthesis production of $^{6+7}$Li or
stellar Li-depletion have been applied;
the uncertainties are undoubtedly dominated by systematic errors.
Additional simulations
of metal-poor stars are important, in order to estimate more
carefully the statistical errors, and also to check whether the
extreme thinness of the Li-plateau found when using 1D models
(Ryan et al. 1999) can be retained with 3D models.
We note that our LTE Li estimate 
is only barely consistent with standard Big Bang nucleosynthesis 
predictions:
if correct, our results imply that the baryon density in the Universe
is $\rho_{\rm B} \approx 1.7 \cdot 10^{-31}$\,g\,cm$^{-3}$
($\eta \approx 2.5 \cdot 10^{-10}$), which is in good agreement
with $^4$He ($Y_{\rm p} = 0.238 \pm 0.002 \pm 0.005$) 
but at variance with both of the contested high and low D measurements 
(e.g. Olive 1999).

A discussion of
the full impact of our results on galactic chemical evolution
is clearly beyond the scope of the present paper. 
Instead, we very briefly point out some implications for the
production of the light elements in the early Galaxy. 
Since Fe presumably suffers from over-ionization,
[Fe] is best represented by [Fe\,{\sc ii}$_{\rm 3D}$] 
which is slightly higher
than [Fe\,{\sc i}$_{\rm 1D}$]. Therefore the slope [Be/H] vs [Fe/H] should
remain essentially unchanged compared with 1D, i.e. close to unity. 
Though OH lines have not been investigated here, we anticipate 
a decrease of the derived O abundance with the 3D simulations, and
thus that the [Be/H] vs [O/H]-relation should be shallower than in 1D.
It would thus be more consistent with a primary origin for Be. 
Likewise we expect the 
[O/Fe] vs [Fe/H]-relation (based on OH-lines) 
to be flatter compared with the most recent
findings (Israelian et al. 1998), but the claimed consistency between
OH, O\,{\sc i} and [O\,{\sc i}] lines can no longer be expected, 
emphasizing the need for a non-LTE study of O based on these
inhomogeneous atmospheres. Furthermore, since Be\,{\sc ii} and B\,{\sc i}
react in opposite ways to the low atmospheric temperatures, the
B/Be ratio may previously have been overestimated.
However, departures from LTE driven by over-ionization are significant 
already in 1D for B (Kiselman \& Carlsson 1996) 
and it seems likely that in 3D the non-LTE effects will be
aggravated, which may well be true also for other elements. 

\begin{acknowledgements} 
It is a pleasure to thank C. Allende Prieto, B. Freytag, B. Gustafsson,
D. Kiselman, H.-G. Ludwig, 
P.E. Nissen and M. Steffen for rewarding discussions.
\end{acknowledgements}


\end{document}